\documentclass[pra,showpacs,showkeys,twocolumn]{revtex4}

\usepackage{amsmath}
\usepackage{amsfonts}            
\usepackage{amssymb}

\usepackage[colorlinks,urlcolor=blue,linkcolor=blue,pagecolor=blue,filecolor=blue,citecolor=blue]{hyperref}

\usepackage{mathrsfs}

\usepackage{eufrak}

\usepackage{verbatim} 

\usepackage{graphicx}
\usepackage{epstopdf}

\usepackage{bm}

\begin{document}

\title{Rotational and rotationless states of weakly-bound molecules}

\author{Mikhail Lemeshko} 

\author{Bretislav Friedrich}

\affiliation{%
Fritz-Haber-Institut der Max-Planck-Gesellschaft, Faradayweg 4-6, D-14195 Berlin, Germany
}%

\date{\today}

\begin{abstract}
By making use of the quantization rule of Raab and Friedrich [P. Raab and H. Friedrich, Phys. Rev. A {\bf 78}, 022707 (2008)], we derive simple and accurate formulae for the number of rotational states supported by a weakly-bound vibrational level of a diatomic molecule and the rotational constants of any such levels up to the threshold, and provide a criterion for determining whether a given weakly-bound vibrational level is rotationless. The results depend solely on the long-range part of the molecular potential and are applicable to halo molecules.
\end{abstract}
\maketitle

Recent experimental studies of ultracold molecules produced by photo- and magneto-association \cite{RevFeshbach}--\cite{RevPhotoassoc} and of other halo molecules~\cite{JensenHalo04} rekindled an interest in the vibrational and rotational structure of these weakly-bound species. In 1970, LeRoy and Bernstein~\cite{LeRoy70} and, independently, Stwalley~\cite{Stwalley70} derived a WKB ``near dissociation equation,'' which related the energies of the vibrational levels close to threshold for dissociation to the long-range behavior of the potential. However, the WKB approximation is invalid for near-threshold molecular states, since, at threshold, the anticlassical limit is reached~\cite{HaraldFri04}. Stimulated especially by the rapid expansion of cold-molecule research, considerable effort has been invested in correcting the WKB approximation to allow for the treatment of weakly-bound vibrational states~\cite{TrostEtchFri98}--\cite{PatrickHarald08}.

In 1972, LeRoy applied the WKB approximation to the rotational structure of weakly-bound molecules~\cite{LeRoy72}, which turned out to be even less accurate than for the vibrational levels. LeRoy concluded that ``only upper and lower bounds could be given, rather than accurate predicted values for unobserved [rotational constants] near [the dissociation limit]."
However, the rotational structure of weakly-bound species is of considerable interest, which is fueled mainly by the current work on magneto-association of ultracold atoms via higher-order Feshbach resonances~\cite{RevFeshbachRudi}, creation of ultracold molecules in highly-excited rotational states~\cite{KnoopGrimm08}, and probing of halo molecules with nonresonant light~\cite{LemFriPRL09}.

Herein, we derive simple and accurate formulae for the number of rotational states supported by a weakly-bound vibrational level of a diatomic molecule and the rotational constants of any such levels up to the threshold, and provide a criterion for determining whether a given weakly-bound vibrational level is rotationless. 

We first introduce the quantization rule of Raab and Friedrich~\cite{HaraldPatrick08},\cite{PatrickHarald08} for rotationless states of diatomic molecules bound by a radial potential which behaves asymptotically as
\begin{equation}
	\label{NoJpot}
	V(r) \overset{r \to \infty}{\sim}  -\frac{C_n}{r^n} \hspace{0.5cm} \text {with}~n>2 
\end{equation}
and thus is a homogeneous function of degree $-n$. Our notation is tailored for molecules and therefore somewhat deviates from that of refs.~\cite{HaraldPatrick08},\cite{PatrickHarald08}. The quantization rule is given by
\begin{equation}
	\label{QuantFuncSimple}
	F(E_\text{b}) = v_\text{th} - v
\end{equation}
where $F(E_\text{b})$ is the quantization function, $v$ is the (integral) vibrational quantum number and $v_\text{th}$ is the nonintegral quantum number that pertains to a level exactly at threshold. The binding energy, $E_\text{b}=D-E_v$, with $D$ the dissociation energy and $E_v$ the energy of the vibrational level $v$, is thus positive, $E_\text{b}>0$. The quantization function $F(E_\text{b})$ was recently derived by Raab and Friedrich for an arbitrary binding energy $E_\text{b}$~\cite{HaraldPatrick08},\cite{PatrickHarald08}. For a homogeneous potential, it is given by:
\begin{equation}
	\label{QuantFunction}
	F(E_\text{b}) = F_\text{th}(\kappa)+F_\text{ip}(\kappa) \Bigl[ F_\text{cr}(\kappa) + F_\text{WKB}(\kappa) \Bigr]
\end{equation}
where the individual terms, herein introduced for convenience, comprise a near-threshold dependence,
\begin{equation}
	\label{CoeffF1}
	F_\text{th}(\kappa) = \frac{2 b \kappa -(p \kappa)^2}{2 \pi \left[1 + (G \kappa)^4 \right]};
\end{equation}
an ``interpolation'' term, which gives a smooth transition between low-$\kappa$ and high-$\kappa$ behavior, 
\begin{equation}
	\label{CoeffF2}
	F_\text{ip}(\kappa) =\frac{(G \kappa)^4}{1+(G \kappa)^4};
\end{equation}
a term which corrects the reflection phase due to the potential of Eq.~(\ref{NoJpot}),
\begin{equation}
	\label{CoeffF3}
	F_\text{cr}(\kappa) =-\frac{1}{2(n-2)} +\frac{u}{2 \pi  \kappa^{1-2/n}};
\end{equation}
and a pure WKB contribution,
\begin{equation}
	\label{CoeffF4}
	F_\text{WKB}(\kappa) =\frac{\kappa^{1-2/n}}{\sqrt{\pi} (n-2)} \frac{\Gamma(\tfrac{1}{2}+\tfrac{1}{n})}{\Gamma(1+\tfrac{1}{n})}
\end{equation}
cf. Eq.~(15) of ref.~\cite{HaraldPatrick08}. In Eqs.~(\ref{QuantFunction})--(\ref{CoeffF4}), the dimensionless wavenumber $\kappa$ is defined by 
 \begin{equation}
	\label{Varkappa}
	\kappa \equiv k \left(\frac{C_n 2 m}{\hbar^2} \right)^\frac{1}{n-2} = E_\text{b}^\frac{1}{2} C_n^\frac{1}{n-2} \left(\frac{2 m}{\hbar^2} \right)^\frac{n}{2 (n-2)}
\end{equation}
with $k = \sqrt{2 m E_\text{b}}/\hbar$ the wavenumber and $m$ the diatomic's reduced mass. The parameters $b$, $p$ and $u$ in eq.~(\ref{QuantFunction}) are defined by 
\begin{equation}
	\label{bParameter}
	b \equiv \sin(\pi y) y^{2y} \frac{\Gamma(1-y)}{\Gamma(1+y)}
\end{equation}
\begin{multline}
	\label{pParameter}
	p^2 \equiv \frac{2 \pi y^{4y+1}}{\Gamma(\tfrac{1}{2} + y)} \frac{\Gamma(1-y)}{\Gamma(1+y)} \\ 
	\times \left[ \frac{2^{2y} \Gamma(\tfrac{1}{2}+2y)}{\Gamma(1+3y)} - \frac{\pi}{\left[\Gamma(1+y) \right]^2 \Gamma(\tfrac{1}{2} -y)}  \right]
\end{multline} 
\begin{equation}
	\label{DParameter}
	u \equiv \sqrt{\pi} \frac{(n+1)}{12 n} \frac{\Gamma(\tfrac{1}{2} - \tfrac{1}{n})}{\Gamma(1 - \tfrac{1}{n})}
\end{equation} 
with $y \equiv 1/(n-2)$ (cf. Eqs.~(10),(12) of ref.~\cite{HaraldPatrick08}). Note that in order to avoid confusion with the rotational constant (defined below), we changed the symbol $B$, used in ref.~\cite{HaraldPatrick08}, to $G$. The adjustable length-parameter $G$, which connects the low-$\kappa$ and high-$\kappa$ behavior, was obtained in ref.~\cite{HaraldPatrick08} for $n=4-7$, and is given below.

\begin{figure}[t]
\includegraphics[width=8.5cm]{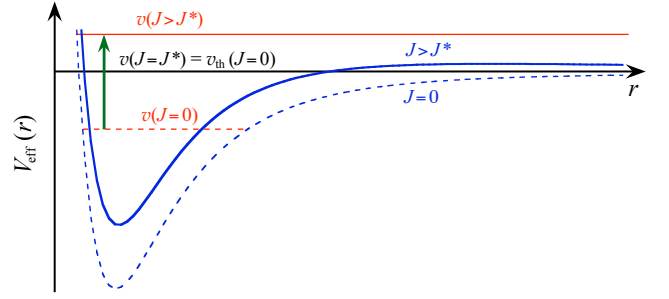}
\caption{\label{fig:eff_pot} (Color online) A schematic illustrating the role of the centrifugal term in the effective potential, Eq. (\ref{RotatingEffPot}); the energy splittings have been exaggerated. Shown is the position of a rotationless vibrational level, $v(J=0)$ (dashed line), as well as the position of the same level when pushed up by the centrifugal term to threshold, $v(J=J^\ast)=v_\text{th}(J=0)$ (full line, at threshold). When the rotational angular momentum $J$ exceeds the critical value $J^\ast$, the centrifugal term pushes the vibrational level above threshold, $v(J>J^\ast)$ (full line, above threshold), thus leading to dissociation.}
\end{figure}

We now turn to the case of a rotating molecule. Molecular rotation adds a repulsive centrifugal term to the attractive inverse-power potential, which gives rise to an effective potential 
\begin{equation}
	\label{RotatingEffPot}
	U(r)  = -\frac{C_n}{r^n} + \frac{\hbar^2}{2 m} \frac{J(J+1)}{r^2}
\end{equation}
with $J$ the rotational angular momentum quantum number. For $J>0$, the centrifugal term pushes the vibrational manifold due to $V(r)$ upwards, thereby reducing the binding energy of the vibrational levels. For each vibrational level, there is a critical value, $J^\ast$, of the angular momentum that pushes the level up to threshold, thereby causing its binding energy to vanish. Hence an angular momentum $J$ in excess of $J^\ast$, $J>J^\ast$, dissociates the molecule. The effect of the centrifugal term leading to dissociation is shown schematically in Fig.~\ref{fig:eff_pot}. 

By making use of Eqs.~(202), (204) and (205) of ref.~\cite{HaraldFri04}, we can express  the threshold quantum number $v_\text{th}$ for the effective potential, Eq.~(\ref{RotatingEffPot}), as a function of $J$,
\begin{multline}
	\label{nthRotating}
	v_\text{th} (J) = \frac{1}{\pi \hbar} \int_{r_\text{in}(0)}^{\tilde{r}} p_0(r) dr  + \frac{2 (2m)^{1/2}}{\pi \hbar (n-2)} \frac{C_n^{1/2}}{\tilde{r}^\frac{n-2}{2}} \\
	- \frac{\phi_\text{in}(0)}{2 \pi} - \frac{1}{4} - \frac{1}{n-2} \sqrt{J (J+1) +\tfrac{1}{4}}
\end{multline}
where $r_\text{in}(0)$ is the inner turning point, $p_0 (r)= \sqrt{2mV_\text{eff}(r)}$ is the  momentum, and $\phi_\text{in}(0)$ is the inner reflection phase, all at threshold (zero energy). The distance $\tilde{r}$ defining the upper limit of the action integral must lie within a region where the WKB approximation is sufficiently accurate and the potential, Eq.~(\ref{RotatingEffPot}), is dominated by the $-C_n/r^n$ term. In such a case the sum of the first two terms of Eq.~(\ref{nthRotating}) is independent of the choice of $\tilde{r}$. Since the dependence of $r_\text{in}(0)$ and $\phi_\text{in}(0)$ on the rotational quantum number is weak-enough to be neglected, the dependence of $v_\text{th}$ on $J$ is given solely by the last term. We note that $v_\text{th} (J)$ maintains the form given by Eq. (\ref{nthRotating}) for any positive $J$, unlike the quantization function and other near-threshold properties of the potential~(\ref{RotatingEffPot}), see Table 5 of ref.~\cite{HaraldFri04}. We now use Eq. (\ref{nthRotating}) to derive a closed-form expression for $J^\ast$, Eq. (\ref{Jast}). 

In the limiting case, when the molecule is transferred from a rotationless state, $J=0$, to a state with a critical angular momentum, $J=J^\ast$, the threshold quantum number, $v_\text{th} (0)$, decreases by:
\begin{multline}
	\label{DeltaNth}
	\Delta v_\text{th} (J^\ast)= v_\text{th} (0) - v_\text{th} (J^\ast)\\
	= \frac{1}{n-2} \left[\sqrt{J^\ast (J^\ast +1) + \frac{1}{4} } - \frac{1}{2} \right]
\end{multline}
Since, at the same time, for a critical angular momentum $J^\ast$, the vibrational state $v$ is pushed up to threshold, its quantum number coincides with the threshold quantum number,
\begin{equation}
	\label{NthJastN}
	v_\text{th} (J^\ast) = v,
\end{equation}
or
\begin{equation}
	\label{NthJastN}
	v_\text{th} (0) - v_\text{th} (J^\ast) = v_\text{th} (0) - v
\end{equation}
see also Fig. \ref{fig:eff_pot}. We thus obtain that the change of the threshold quantum number, $\Delta v_\text{th} (J^\ast)=v_\text{th} (0) - v$, is nothing else than the quantization function, Eq.~(\ref{QuantFunction}), for the rotationless potential $V(r)$, Eq.~(\ref{NoJpot}):
\begin{equation}
	\label{ConditionJast}
	\Delta v_\text{th} (J^\ast) = F(E_\text{b})
\end{equation}
By combining Eqs.~(\ref{DeltaNth}) and~(\ref{ConditionJast}) we obtain a closed-form expression for $J^\ast$:
\begin{equation}
	\label{Jast}
	J^\ast = F(E_\text{b}) (n - 2)
\end{equation}
This simple formula renders $J^\ast$ quite accurately, as exemplified in Table~\ref{table:ResultsRb2}. For instance, for the three last vibrational states of $^{85}$Rb$_2$, Eq.~(\ref{Jast}) yields values of $J^\ast$ which are in a very good agreement with the essentially exact values $J^\ast_\text{exact}$, obtained by solving the Schr\"odinger equation with the potential energy curves taken from refs.~\cite{Seto00},\cite{vanKempen02}. 

We note that  Gao used the angular-momentum-insensitive quantum defect theory to obtain the number of rotational states which are supported by weakly-bound vibrational levels~\cite{Gao04}. The theory, which, in general, requires evaluating the quantum defect, gives the upper bound on the number of rotational states supported by the least-bound vibrational level, $J_\text{max} = n-2$. In our work, $J_\text{max}$ is simply given by the integer part of $J^\ast$, $J_\text{max}=\text{Int} [J^\ast]$. Hence Gao's result is seen to be a limiting case of Eq.~(\ref{Jast}) for $v_\text{th}-v = 1$. 

Neglecting any coupling of the molecular rotation, we can estimate the rotational constant, $B$, from the rotational energy, $BJ^\ast(J^\ast+1)$, required to promote the vibrational level bound by $E_\text{b}$ to threshold
\begin{equation}
	\label{BrotConst}
	B = \frac{E_\text{b}}{J^\ast(J^\ast+1)}
\end{equation}
The values of the rotational constant $B$ obtained from Eq.~(\ref{BrotConst}) for $^{85}$Rb$_2$ are listed in Table~\ref{table:ResultsRb2} together with the essentially exact values, $B_\text{exact}$. The latter were calculated from
\begin{equation}
	\label{Bexact}
	B_\text{exact} = \langle v \vert \frac{\hbar^2}{2 m r^2} \vert v \rangle 
\end{equation}
with the vibrational wavefunctions obtained from a numerical solution of the Schr\"odinger equation for the potential of refs.~\cite{Seto00},\cite{vanKempen02}. 

\begin{table}[t]
\caption{Comparison of the critical angular momenta $J^\ast$ and rotational constants $B$ obtained for the three least-bound states of the $^{85}$Rb$_2$ dimer in different approximations; $E_\text{b}$ and $B$ are given in 10$^{-4}$ cm$^{-1}$. See also Table \ref{table:Assumptions} and text.}
\vspace{0.5cm}
\label{table:ResultsRb2}
\begin{tabular}{ c c c c c  c  c c  c   c c  c  c}
 \hline
\hline
$v$ && $E_\text{b}$ && $J^\ast$  & $J^\ast_\text{exact}$ & $J^\ast_\text{iLB}$ & $J^\ast_\text{LB}$ && $B$  & $B_\text{exact}$ & $B_\text{iLB}$ & $B_\text{LB}$  \\[3pt]
\hline
123 && 0.08 && 0.22 & 0.22 & 0.43 & 0.44 && 0.30 &  0.41 & 0.13 & 0.13 \\[3pt]
122 && 97.9 && 4.25 & 4.25 & 4.25 & 4.71 && 4.39 &  4.73 & 4.39 & 3.65 \\[3pt]
121 && 630.6 && 8.28 & 8.48  & 8.28 & 8.76 && 8.21 &  8.68 &  8.21 &  7.38 \\[3pt]
  \hline
 \hline
\end{tabular}
\end{table}

From Table~\ref{table:ResultsRb2} one can see that for the least-bound state, with $v=123$, whose binding energy is only $8\times10^{-6}$~cm$^{-1}$, Eq.~(\ref{BrotConst}) yields a value of $B$ which is by about 25\% smaller than the exact one. However, for the two lower vibrational levels, $v=121$ and $v=122$, each of which supports several rotational states, Eq.~(\ref{BrotConst}) gives a value of $B$ which is less than ten percent off the exact one. 

\begin{table}[t]
\caption{Terms of the quantization function of Raab and Friedrich (RF), Eq.~(\ref{QuantFunction}), inherent to the LB and the iLB approximations. Also shown are the ranges of the reduced wavenumber $\kappa$ wherein the approximations apply. See text.}
\vspace{0.5cm}
\label{table:Assumptions}
\begin{tabular}{ c c c c c  c c c c}
 \hline
\hline
RF &&&& iLB  &&&& LB   \\[3pt]
\hline
All terms &&&& \begin{tabular}{c} $F_\text{th}=0$ \\  $F_\text{ip}=1$  \end{tabular} &&&& \begin{tabular}{c} $F_\text{th}=0$ \\  $F_\text{ip}=1$ \\  $F_\text{cr}=0$  \end{tabular}  \\[3pt]
All $\kappa$ &&&& $\kappa \approx 1$  &&&& $\kappa \gg 1$   \\[3pt]
  \hline
 \hline
\end{tabular}
\end{table}

Table \ref{table:ResultsRb2} also lists values of $J^\ast$ and $B$ obtained from the LeRoy-Bernstein (LB) and the improved LeRoy-Bernstein (iLB) approximations. The LB approximation results when, in the quantization function of Eq.~(\ref{QuantFunction}), $F_\text{th}(\kappa)$ and $F_\text{cr}(\kappa)$ are neglected,  and $F_\text{ip}(\kappa)$ is set to $\approx 1$. What remains is the WKB term which gives the semiclassical LeRoy-Bernstein (LB) quantization condition~\cite{LeRoy70}.
The iLB approximation comprises $F_\text{cr}(\kappa)$ and $F_\text{WKB}(\kappa)$ but neglects $F_\text{th}(\kappa)$ and $F_\text{ip}(\kappa)$, thereby accounting for short-range deviations of the true potential from $V(r)$ of Eq.~(\ref{NoJpot}), see refs.~\cite{Jelassi0608},\cite{HaraldPatrick08}. The assumptions about the various terms of Eq.~(\ref{QuantFunction}) inherent to the approximations are listed in Table.~\ref{table:Assumptions}.

From Table \ref{table:ResultsRb2}, we  see that in the case of the most weakly-bound state, $v=123$, both the LB and iLB approximations fail badly. The reason is the omission of the $F_\text{th}(\kappa)$ term which actually determines the near-threshold behavior of $F(E_\text{b})$, see Table \ref{table:Assumptions}. As a matter of fact, if for $v=123$ only the $F_\text{th}(\kappa)$ term were nonzero, we would obtain $J^\ast_\text{th}=0.21$ and $B_\text{th}=0.31$. 
On the other hand, in the case of the $v=121$ and $v=122$ levels, which are relatively far from threshold, neglecting the $F_\text{th}(\kappa)$ term is better justified. Indeed, the results of the iLB approximation are in good agreement with the accurate ones. However, since the $v=121$ and $v=122$ levels are still well within the anticlassical region of binding energies, the purely semiclassical LB approximation remains inaccurate.

In the context of the present work, one can easily answer the question as to which of the weakly bound vibrational levels are rotationless, see also ref.~\cite{LemFriPRL09}. Indeed, the value of the binding energy $E_\text{b}$ required in order for the first rotational state to be supported by the potential is given by Eq.~(\ref{Jast}) with $J^\ast=1$. By combining this requirement with Eqs.~(\ref{QuantFunction})--(\ref{Varkappa}), we obtain at once a criterion for a vibrational level to be rotationless, namely, when its binding energy satisfies the condition
\begin{equation}
	\label{CriterionEb}
	E_\text{b} < d_n \left( \frac{\hbar^2}{m~C_n^\frac{2}{n}} \right)^\frac{n}{n-2}
\end{equation}

The parameter $d_n$ for an inverse-power potential, Eq.~(\ref{NoJpot}), depends solely on the power $n$. It can be obtained by solving Eq.~(\ref{Jast}) with $J^\ast=1$ numerically. The values of $d_n$ are listed in Table~\ref{table:parameters} for $n=4-7$. We verified the validity of the criterion, Eq.~(\ref{CriterionEb}), for the case of $n=6$ by solving the Schr\"odinger equation for the Lennard-Jones (12,6) potential, and found that this agreed with the value of $d_6$ given in Table \ref{table:parameters} within 1\%. Also, we found that Eq.~(\ref{CriterionEb}) provides correct predictions of the rotational structure of the last bound states of the Rb$_2$ and KRb dimers, which we recently investigated~\cite{LemFriPRL09} using the potential curves of refs.~\cite{Seto00}--\cite{Pashov07}.

Within the LB and iLB approximations, the parameter $d_n$ can be evaluated analytically:
\begin{equation}
	\label{dalphaLB}
	d_n^\text{LB} = \left[  \sqrt{\frac{\pi}{2}} \frac{\Gamma(1+\tfrac{1}{n})}{\Gamma(\tfrac{1}{2}+\tfrac{1}{n})} \right]^\frac{2n}{n-2}
\end{equation}
\begin{multline}
	\label{dalphaILB}
	d_n^\text{iLB} = \Biggl[ \frac{3}{4} \sqrt{\frac{\pi}{2}} \frac{\Gamma(1+\tfrac{1}{n})}{\Gamma(\tfrac{1}{2}+\tfrac{1}{n})} \\
	\times \Biggl(1 + \sqrt{1 - u \frac{8 (n-2)}{9\pi^\frac{3}{2}} \frac{\Gamma(\tfrac{1}{2}+\tfrac{1}{n})}{\Gamma(1+\tfrac{1}{n})}} \Biggr) \Biggr]^\frac{2n}{n-2}
\end{multline}
The values of $d_n^\text{LB}$ and $d_n^\text{iLB}$ for different $n$ are listed in Table~\ref{table:parameters}.

\begin{table}[t]
\caption{Values of the parameter $d_n$ appearing in the criterion, Eq. (\ref{CriterionEb}), which determines whether a vibrational level is rotationless. We also present the values from ref.~\cite{HaraldPatrick08} of the adjustable-length parameter $G$. See text.}
\vspace{0.5cm}
\label{table:parameters}
\begin{tabular}{ c c  c  c  c }
 \hline
\hline
$n$ & 4 & 5 & 6 & 7 \\[3pt]
\hline
$d_n$ &  2.8974 & 1.9738 & 1.6014  & 1.3961 \\[3pt]
$d_n^{iLB}$ & 2.8833 & 1.9289  &  1.5233 & 1.2835\\[3pt]
$d_n^{LB}$ & 0.73857 & 0.66932 & 0.63308 & 0.61028 \\[3pt]
\hline
$G$ & 2.3528 & 1.3035 & 0.93323 & 0.73446 \\
  \hline
 \hline
\end{tabular}
\end{table}

One can see that the $d_n^\text{iLB}$ parameters come close to the accurate ones, especially for small $n$, whereas the $d_n^\text{LB}$ parameters are quite off. This is because at the binding energies large-enough for the molecule to support rotational states, the wavenumber $\kappa \approx 1$, and thus neither $\kappa \ll 1$ (near-threshold) nor $\kappa \gg 1$ (WKB) limits apply, see also Table \ref{table:Assumptions}. 

In summary, we undertook a study of the rotational structure of weakly-bound molecules, in which we relied on the
quantization rule of Raab and Friedrich~\cite{HaraldPatrick08},\cite{PatrickHarald08} valid from the classical to the
``anticlassical" limit. We found analytic expressions for the critical
value of the angular momentum that leads to rotational predissociation; the number of rotational states supported by a given vibrational level; the rotational constants of weakly-bound molecules in a given vibrational state; and a criterion for rotationlessness of a vibrational level. All of the above was found to depend just on the long-range potential and to check well against essentially exact calculations done by numerically solving Schr\"odinger's equation for the very well known potentials of Rb$_2$ and KRb. 

 We dedicate this paper to Rostislav Vedrinskii on the occasion of his 70th birthday. We are grateful to Gerard Meijer for encouragement and support, and to Patrick Raab for discussions.


\begin{thebibliography}{99}

	\bibitem{RevFeshbach}
	T.~K\"{o}hler, K.~Goral, P.~S.~Julienne, Rev.~Mod.~Phys. \textbf{78}, 1311 (2006).

	\bibitem{RevFeshbachRudi}
	C.~Chin, R.~Grimm, P.~Julienne, E.~Tiesinga, arXiv:0812.1496 (2008).
	
	\bibitem{RevPhotoassoc}
	K.~M.~Jones, E.~Tiesinga, P.~D.~Lett, P.~S.~Julienne, Rev.~Mod.~Phys. \textbf{78}, 483 (2006).

	\bibitem{JensenHalo04}
	A.~S.~Jensen, K.~Riisager, D.~V.~Fedorov, E.~Garrido, Rev.~Mod.~Phys. \textbf{76}, 215 (2004).
	
	\bibitem{LeRoy70}
	R.~J.~LeRoy, R.~B.~Bernstein, J.~Chem.~Phys.~\textbf{52}, 3869 (1970).

	\bibitem{Stwalley70}
	W.~C.~Stwalley, Chem.~Phys.~Lett. \textbf{6}, 241 (1970).
	
	\bibitem{HaraldFri04}
	H.~Friedrich, J.~Trost, Phys.~Rep.~\textbf{397}, 359 (2004); \textit{ibid}~\textbf{451}, 234 (2007).

	\bibitem{TrostEtchFri98}
	J.~Trost, C.~Eltschka, H.~Friedrich, J.~Phys.~B \textbf{31}, 361 (1998).

	\bibitem{Boisseau98}
	C.~Boisseau, E.~Audouard, J.~Vigu\'e, Eur.~Phys.~Lett. \textbf{41}, 349 (1998).

	\bibitem{Boisseau00}
	C.~Boisseau, E.~Audouard, J.~Vigu\'e, V.~V.~Flambaum, Eur.~Phys.~J.~D \textbf{12}, 199 (2000).
	
	\bibitem{EtchMorFri00}
	C.~Eltschka, M.~J.~Moritz, H.~Friedrich, J.~Phys.~B \textbf{33}, 4033 (2000).
	
	\bibitem{Comparat04}
	D.~Comparat, J.~Chem.~Phys.~\textbf{120}, 1318 (2004).
	
	\bibitem{Jelassi0608}
	H.~Jelassi, B.~Viaris~de~Lesegno, L.~Pruvost, Phys.~Rev.~A \textbf{73}, 032501 (2006); \textit{ibid}~\textbf{77}, 062515 (2008).
	
	\bibitem{HaraldPatrick08}
	H.~Friedrich, P.~Raab, Phys.~Rev.~A \textbf{77}, 012703 (2008).

	\bibitem{PatrickHarald08}
	P.~Raab, H.~Friedrich, Phys.~Rev.~A \textbf{78}, 022707 (2008).

	\bibitem{LeRoy72}
	R.~J.~LeRoy, Can.~J.~Phys.~\textbf{50}, 953 (1972).

	\bibitem{KnoopGrimm08}
	S.~Knoop, M.~Mark, F.~Ferlaino, J.~G.~Danzl, T.~Kraemer, H.~C.~N\"agerl, R.~Grimm, Phys.~Rev.~Lett. \textbf{100}, 083002 (2008).	
	
	\bibitem{LemFriPRL09}
	M.~Lemeshko, B.~Friedrich, submitted (2009); arXiv:0903.0811.
	


	\bibitem{MorEtchFri01}
	M.~Moritz, C.~Eltschka, H.~Friedrich, Phys.~Rev.~A \textbf{63}, 042102 (2001); \textit{ibid}~\textbf{64}, 022101 (2001).
		

	\bibitem{Seto00} 
	J.~Y.~Seto, R.~J.~Le~Roy, J.~Verg\`es, C.~Amiot, J.~Chem.~Phys. {\bf 113}, 3067 (2000).
	
	\bibitem{vanKempen02} 
	E.~G.~M.~van~Kempen, S.~J.~J.~M.~F.~Kokkelmans, Phys.~Rev.~Lett. {\bf 88}, 093201 (2002).	


	\bibitem{Gao04}
	B.~Gao, Eur.~Phys.~J.~D \textbf{31}, 283 (2004).

	
	\bibitem{Pashov07} 
	A.~Pashov, O.~Docenko, M.~Tamanis, R.~Ferber, H.~Kn\"ockel, E.~Tiemann, Phys.~Rev.~A {\bf 76}, 022511 (2007).
	

	


\end{thebibliography}
\end{document}